%% file: main.tex
\PassOptionsToPackage{dvipsnames}{xcolor}

\documentclass[twocolumn,twocolappendix]{aastex701}
\input{./vectors}
\input{./units}
\input{./symbols}
\input{./nuclides}

\input{./formatting}

\input{./derivatives}

\input{./concepts}

\input{./code}


\input{./abbrev}

\usepackage{xcolor}

\shorttitle{Mixing Conditions and Light Odd-Z Elements}
\shortauthors{J. Issa et al.}

\graphicspath{{./}}

\begin{document}

\title{3D Macro Physics and Light Odd-Z Element Production in O-C Shell Mergers: Implications for \potassium[40] production and radiogenic heating inventories of rocky exoplanets}

\author[0000-0002-1283-6636,gname=Joshua,sname=Issa]{Joshua Issa}
\affiliation{Astronomy Research Centre, Department of Physics \& Astronomy, University of Victoria, Victoria, BC, V8W 2Y2, Canada}
\affiliation{NuGrid Collaboration, \url{http://nugridstars.org}}
\email[show]{joshuaissa@uvic.ca}  

\correspondingauthor{Joshua Issa}

\author[0000-0001-8087-9278,gname=Falk,sname=Herwig]{Falk Herwig}
\affiliation{Astronomy Research Centre, Department of Physics \& Astronomy, University of Victoria, Victoria, BC, V8W 2Y2, Canada}
\affiliation{NuGrid Collaboration, \url{http://nugridstars.org}}
\email{fherwig@uvic.ca}

\author[0000-0003-0000-125X,gname=Stephen,gname=J., sname=Mojzsis]{Stephen J. Mojzsis}
\affiliation{Geoastronomy Research Group, Bayerisches Geoinstitut, Universität Bayreuth, Universitätsstraße 30, 95447 Bayreuth, Germany}
\affiliation{HUN-REN Research Center for Astronomy and Earth Sciences (CSFK), 1121 Budapest, Konkoly Thege Miklós út 15-17, Hungary}
\affiliation{CSFK, MTA Centre of Excellence, Budapest, Konkoly Thege Miklós út 15-17., H-1121, Hungary}
\email{stephen.mojzsis@uni-bayreuth.de}

\author[0000-0002-9048-6010,gname=Marco,sname=Pignatari]{Marco Pignatari}
\affiliation{HUN-REN Research Center for Astronomy and Earth Sciences (CSFK), 1121 Budapest, Konkoly Thege Miklós út 15-17, Hungary}
\affiliation{CSFK, MTA Centre of Excellence, Budapest, Konkoly Thege Miklós út 15-17., H-1121, Hungary}
\affiliation{Geoastronomy Research Group, Bayerisches Geoinstitut, Universität Bayreuth, Universitätsstraße 30, 95447 Bayreuth, Germany}
\affiliation{NuGrid Collaboration, \url{http://nugridstars.org}}
\email{mpignatari@gmail.com}

\received{September 23, 2025}
\revised{December 8, 2025}
\accepted{December 8, 2025}

\submitjournal{ApJ}

\begin{abstract}
The light odd-Z elements \phosphorus[], \chlorine[], \potassium[], and \scandium[] are underproduced
in galactic chemical evolution models compared to spectroscopic observations of stars in the Milky
Way. The most promising solution to this puzzle is that some massive stars experience
\oxygen[]-\carbon[] shell mergers boosting their yields through dynamic, convective-reactive
nucleosynthesis. We report how convective macro physics based on 3D $4\pi$ hydrodynamic simulations
impacts production in the \oxygen[] shell by post-processing the $\mzams=\unit{15}{\Msun}$ $Z=0.02$
model from the NuGrid dataset. We explore a mixing downturn, boosted velocities, reduced ingestion
rate, and convective quenching. Across 24 mixing cases, the pre-explosive yields for
[\phosphorus[]/\iron[]], [\chlorine[]/\iron[]], [\potassium[]/\iron[]], and [\scandium[]/\iron[]]
are modified by $\unit{[-0.33,0.23]}{\dex}$, $\unit{[-0.84,0.64]}{\dex}$,
$\unit{[-0.78,1.48]}{\dex}$, and ${[-0.36,1.29]}{\dex}$, respectively. Cases with a convective
downturn with the fastest ingestion rate have the largest enhancement, and production is
non-monotonic with boosted velocities. Which reactions are most important for the
convective-reactive element production pathways depends on the mixing case. We parameterize production of
\potassium[40] ($t_{1/2} = \unit{1.248}{\Gyr}$), an important radiogenic heat source for younger
(\unit{2{-}3}{\Gyr}) rocky planets and find a yield variation exceeding three orders of magnitude.
This range of initial abundances for \potassium[40] implies the early geodynamic behaviour of
silicate mantles in rocky planets can differ greatly from that of Earth. These results underscore
the importance of investigating the 3D macro physics of shell merger convection through hydrodynamic
simulations to develop a predictive understanding of the origin and variability of the light odd-Z
elements and the \potassium[40]/\potassium[] ratio in planet host stars. 
\end{abstract}


\section{Introduction}\label{sec:intro} 

The radiogenic isotope \potassium[40] with a half-life of \unit{1.248}{\Giga \yr} is the dominant heat-producing nuclide for rocky
exoplanets in the first few billion years of their geologic histories \citep[e.g.][]{oneillEffectGalacticChemical2020}, with Earth as the baseline example \citep{korenagaPlateTectonicsFlood2008}.
Of the long-lived radionuclides present in silcate mantles of rocky planets, \potassium[40]-driven radiogenic heating contributes to mantle convection that feeds into volcanism, the primary mechanism for transferring volatiles from planetary interiors to atmospheres.
Variations in initial \potassium[40] inventories directly affect the duration and intensity of volcanic degassing \citep{noack:14}, particularly during the critical first $2{-}3$ billion years when secondary atmosphere formation occured when \potassium[40] was still a dominant heat-producer \citep{fischer:20}. 

Therefore, any model of the geodynamic evolution of rocky exoplanets requires knowledge of the initial \potassium[40] inventories.
However, \potassium[] belongs to a group of elements with a particularly uncertain production history. 
A long-standing problem has been that the light odd-Z elements \phosphorus[], \chlorine[], \potassium[], and \scandium[] are produced in insufficient quantities in galactic chemical evolution (GCE) models. 
Indeed, GCE models predict that the majority contribution for these elements comes from core-collapse supernovae (CCSNe), but they cannot match observed stellar abundances \citep[e.g.\,][]{mishenina:17, prantzos:18, kobayashiOriginElementsCarbon2020}. 
A recent realization has been that the light odd-Z elements are significantly produced during \oxygen[]-\carbon[] shell merger in massive stars prior to a CCSN at all metallicities \citep{rauscherNucleosynthesisMassiveStars2002,ritterConvectivereactiveNucleosynthesisSc2018,robertiOccurrenceImpactCarbonOxygen2025}, and GCE models including this effect parametrically can match stellar abundance observations.

Current \potassium[] and by extension \potassium[40] predictions from O-C shell mergers remain uncertain. 
A key source of uncertainty is the impact of macro physics of convection during the dynamic nucleosynthesis, possibly with feedback, as demonstrated recently for the \ppr-production in
O-C shell mergers \citep{issaImpact3DMacro2025}.  A \potassium[40] production variability similar to
that seen for \ppr\ abundance predictions taking into account different scenarios of macro physics
of convection based on stellar hydrodynamics simulations, could lead to divergent planetary
atmospheric evolution pathways. Differences in composition and thickness of planet atmospheres
depend on both the thermal state driving outgassing efficiency and the pressure-temperature
conditions controlling volatile speciation \citep{ortenzi:20}, creating potentially observable
signatures in CO$_2$/H$_2$O ratios and atmospheric oxidation states that future missions like the
Habitable Worlds Observatory could detect \citep{liggins:22}. A predictive, quantitative
understanding of \potassium[40] production would remove a key roadblock in our ability to model the
thermal evolution of silicate-metal exoplanets \citep{frankRadiogenicHeatingEvolution2014}.

Not surprisingly, three-dimensional (3D) hydrodynamic simulations of the \oxygen[] shell where
the nucleosynthesis occurs show very different mixing conditions compared to one-dimensional (1D)
predictions using mixing length theory (MLT)
\citep{meakinActiveCarbonOxygen2006,meakinTurbulentConvectionStellar2007,jonesIdealizedHydrodynamicSimulations2017,yoshidaOneTwoThreedimensional2019,
andrassy3DHydrodynamicSimulations2020,yadavLargescaleMixingViolent2020a,yoshidaThreedimensionalHydrodynamicsSimulation2021,fieldsThreedimensionalHydrodynamicSimulations2021,
rizzutiShellMergersLate2024a}.
These simulations have shown that the radial mixing efficiency
profile has a downturn as the convective boundary is approached, which is not captured in 1D models
which predict the mixing should be nearly constant across the convective shell
\citep{meakinActiveCarbonOxygen2006,meakinTurbulentConvectionStellar2007,jonesIdealizedHydrodynamicSimulations2017}.
They also predict that the convective velocities could be larger by up to a factor of $30$ or more
than what is predicted by 1D models \citep{jonesIdealizedHydrodynamicSimulations2017,
andrassy3DHydrodynamicSimulations2020,rizzutiShellMergersLate2024a}. Finally, convective quenching
due to energy feedback from the ingestion of fresh fuel has been demonstrated in 3D simulations of
H-ingestion in post-AGB (asymptotic giant branch; AGB) stars
\citep{herwigGLOBALNONSPHERICALOSCILLATIONS2014}, and  the ingestion of \carbon[]-shell material
into the \oxygen[] shell may play a similar restricting role in \oxygen[]-\carbon[] shell mergers
\citep{andrassy3DHydrodynamicSimulations2020}.

These mixing conditions directly affect the nucleosynthesis as the \oxygen[] shell is a
convective-reactive environment where the timescales for mixing and burning are comparable
\citep{ritterConvectivereactiveNucleosynthesisSc2018,yadavLargescaleMixingViolent2020a,rizzutiShellMergersLate2024a}.
\cite{rizzutiShellMergersLate2024a} have shown in their 3D simulation of an \oxygen[]-\carbon[]
shell merger that all isotopes have significantly different results compared to 1D simulations.

Here we consider how the isotopes \phosphorus[31], \chlorine[35,37], \potassium[39,40,41], and
\scandium[45] are produced in the \oxygen[] shell and how it is impacted by the mixing conditions.
This extends the work done by \cite{issaImpact3DMacro2025} that analyzed how the \pnucn{} are
affected by the mixing conditions during the shell merger. We analyze the production of these
nuclides, with special attention to K and its isotopes. While stellar spectral data of the host star
can be used to infer plausible major element compositions of rocky exoplanets
\citep[e.g.][]{spaargarenPlausibleConstraintsRange2023}, only GCE models can provide likely relative
abundances of the radioactive nuclides (\potassium[40], \thorium[232], \uranium[235-238]), tied to
the age of that system, that can be used to infer heat production in the interiors of rocky
exoplanets \citep{frankRadiogenicHeatingEvolution2014}. While indeed for Th it is possible to
directly observe \thorium[232] in stars in the Milky Way disk, current spectroscopic observation
uncertainties still hinder precise derivation of its abundances \citep[e.g.,][and references
therein]{botelho:19, mishenina:22}.

This paper is structured as follows. In \Sect{methods}, we describe our methodology, including the
initial conditions from the NuGrid stellar model, the post-processed models with different mixing
scenarios, and our approach for estimating final yields and reaction contributions. \Sect{results}
presents our main results, focusing on the nucleosynthesis of light odd-Z isotopes and how different
mixing conditions impact their production. In \Sect{k40_planets}, we discuss the production of
\potassium[40] and its implications for heterogeneous pollution of the interstellar medium and
radiogenic heating in rocky exoplanets. Finally, \Sect{discussion} provides our discussion and
conclusions, highlighting the importance of mixing conditions for understanding the origin of light
odd-Z elements, their impact on galactic chemical evolution models and radiogenic heating in rocky
exoplanets.

\section{Methodology} \label{sec:methods}

\subsection{The O-C shell merger model}
\label{sec:methods_model}
The radial stratification for the \oxygen[] shell multizone nucleosynthesis simulations in this paper are taken from the non-rotating, 1D \unit{15}{\Msun} $Z=0.02$ massive star model from the NuGrid data set \citep{ritterNuGridStellarData2018} calculated with the \MESA{} stellar evolution code \citep{paxtonMODULESEXPERIMENTSLAR2010} and post-processed with the multizone NuGrid code \mppnp{} \citep{pignatariNuGridStellarData2016}.
This model does not have exponential-diffusive convective boundary mixing \citep{freytagHydrodynamicalModelsStellar1996,herwigEvolutionAGBStars2000} after the
cessation of core He burning \citep[which was explored in][]{davis:2018}. The evolution of the
convection zones is shown in a Kippenhahn diagram (Figure 1) in \cite{issaImpact3DMacro2025}. The
light odd-Z elements are mildly produced in the first convective \oxygen[] shell, which extends from
\unit{1.55}{\Msun} to \unit{1.95}{\Msun} during a time interval $\log_{10}(t -
t_{\mathrm{end}})/\mathrm{yr} = -1.76$ to $-2.16$, but the \oxygen[]-\carbon[] shell merger at
$\log_{10}(t - t_{\mathrm{end}})/\mathrm{yr} = -3.85$ is the dominant source of the elements \citep{ritterConvectivereactiveNucleosynthesisSc2018}. 

Table \ref{tab:merger_properties} summarizes the NuGrid model properties with \oxygen[]-\carbon[] shell mergers from \cite{ritterNuGridStellarData2018}. 
The $12$, $15$, and $\unit{20}{\Msun}$ models show clear differences in the size of the merged shells $\Delta m$, duration of the merger $\Delta t$, and overproduction factors \OP{} for \potassium[39-41].  
However, both \unit{15}{\Msun} models with metallicities $Z=0.01$ and $Z=0.02$ are quite similar in all these properties, including in \OP{}, although the $Z=0.02$ model produces three times more \potassium[39] by the end of the merger because of a higher initial mass fraction. 
This would suggest that initial stellar mass is more determinate of the merger and its nucleosynthesis than its metallicity.
In fact, \cite{robertiOccurrenceImpactCarbonOxygen2025} shows that models with \oxygen[]-\carbon[] shell mergers do not have a general trend in [\potassium[]/\magnesium[]] across metallicities of $Z=10^{-9}{-}10^{-2}$.

\begin{table*}[ht]
\centering
\begin{tabular}{cccccccccc}
\toprule
\mzams & $Z_{\mathrm{ini}}$ & $\Delta m$ & $\Delta t$ & $\langle X(\potassium[39])
\rangle_{\mathrm{fin}}$ & \OP(\potassium[39]) & $\langle X(\potassium[40]) \rangle_{\mathrm{fin}}$ &
\OP(\potassium[40]) & $\langle X(\potassium[41]) \rangle_{\mathrm{fin}}$ & \OP(\potassium[41]) \\
\toprule
\unit{12}{\Msun} & 0.01 & \unit{0.6}{\Msun} & \unit{0.029}{\hour} & \natlog{2.48}{-4} & 1.531 &
\natlog{6.72}{-6} & 1.362 & \natlog{1.49}{-6} & 0.487 \\
\unit{15}{\Msun} & 0.01 & \unit{1.52}{\Msun} & \unit{0.74}{\hour} & \natlog{3.01}{-4} & 1.98 &
\natlog{3.24}{-5} & 1.934 & \natlog{2.71}{-5} & 1.63 \\
\unit{20}{\Msun} & 0.01 & \unit{3.6}{\Msun} & \unit{5.43}{\hour} & \natlog{2.28}{-4} & 1.841 &
\natlog{3.56}{-6} & 0.895 & \natlog{7.01}{-7} & 0.004 \\
\unit{15}{\Msun} & 0.02 & \unit{1.43}{\Msun} & \unit{1.252}{\hour} & \natlog{9.43}{-4} & 2.028 &
\natlog{5.85}{-5} & 1.965 & \natlog{2.38}{-5} & 1.305 \\
\toprule
\end{tabular}
\caption{Properties of the merger and production of K for the NuGrid models with O-C shell mergers.
The averaged mass fractions are within the maximum extent of the merger region and
$\OP=\log_{10}(\langle X_\mathrm{fin}\rangle/\langle X_\mathrm{ini}\rangle)$ where $\langle
X_\mathrm{ini}\rangle$ and $\langle X_\mathrm{fin}\rangle$ are taken at the start and end of the
merger. Note that the models have different initial mass fractions of \potassium[39-41].}
\label{tab:merger_properties}
\end{table*}

\subsection{Post-Processed Models}
We use the post-processed models from \cite{issaImpact3DMacro2025} of the \unit{15}{\Msun} $Z=0.02$ NuGrid model that implement insights from 3D hydrodynamic simulations to motivate mixing scenarios expressed in terms of mixing efficiency profiles that are different from the MLT 1D predictions.
We focus solely on the \unit{15}{\Msun} $Z=0.02$ model for this exploratory study as it has the largest production of the \pnucn{} \citep{ritterConvectivereactiveNucleosynthesisSc2018} and \potassium[39-41] (\Tab{merger_properties}) for the merger models.
These simulations use mass fractions from just before merger onset at $\log_{10}(t-t_{\mathrm{end}})/\yr=-3.845$ and a constant stellar structure, including mixing profile, from the onset at $\log_{10}(t-t_{\mathrm{end}})/\yr=-3.856$.
This is run for \unit{110}{\second}, the same time between these moments, which corresponds to at least $2-3$ convective turnovers which implies a quasi-equilibrium state.
All nucleosynthesis is captured with a network of 1470 isotopes.

The MLT mixing case with a mixing efficiency profile essentially constant across the \oxygen[] shell as taken from \cite{ritterNuGridStellarData2018}. 
3D simulations, however, predict a mixing efficiency downturn as the convective boundary is approached and an increased mixing efficiency compared to MLT \citep{meakinTurbulentConvectionStellar2007, jonesIdealizedHydrodynamicSimulations2017,andrassy3DHydrodynamicSimulations2020, fieldsThreedimensionalHydrodynamicSimulations2021, rizzutiShellMergersLate2024a}, to some degree depending on the uncertain hydrodynamic response to the ingestion and burning of additional nuclear fuel. 
Therefore, we implement the 3D-inspired mixing profile from \cite{jonesIdealizedHydrodynamicSimulations2017} that features a convective downturn, and boost the MLT velocities by factors of $1$, $3$, $10$, and $50$.
The MLT and 3D-inspired cases are calculated with ingestion rates of $\unit{\natlog{4}{-5}}{\Msun\second^{-1}}$, $\unit{\natlog{4}{-4}}{\Msun\second^{-1}}$, $\unit{\natlog{4}{-3}}{\Msun\second^{-1}}$, which are consistent with those adopted in the 3D simulations of \cite{andrassy3DHydrodynamicSimulations2020}.
The extent of the \carbon[] shell is \unit{0.8}{\Msun}, so for a post-processing simulation time of \unit{110}{\second}, the maximum ingestion rate would be $\unit{\natlog{7}{-3}}{\Msun\second^{-1}}$ like \cite{ritterConvectivereactiveNucleosynthesisSc2018}. 
Our fastest ingestion rate of $\unit{\natlog{4}{-3}}{\Msun\second^{-1}}$ reflects a full merger and a case without ingestion is also included for reference. 
Quenched mixing cases are also considered by implementing a dip in the mixing efficiency profile.
This may arise due to energy feedback in the middle of the shell at \unit{4.95}{\Mm} and partial merging of the shells at \unit{7.5}{\Mm} as 3D simulations suggest \citep{andrassy3DHydrodynamicSimulations2020}.
Two quenching strengths expressed as dips in the profile to $\unit{10^{14}}{\cm^2\second^{-1}}$ and $\unit{10^{13}}{\cm^2\second^{-1}}$ are considered.
The quenched mixing cases are calculated with an ingestion rate of $\unit{\natlog{4}{-3}}{\Msun\second^{-1}}$ only.
There are 4 MLT, 16 3D-inspired, and 4 quenched mixing cases for a total of 24 simulations.
This reflects both our limited knowledge of and the intrinsically varying conditions in real 3D stars. 
The \mppnp{} code treats mixing as a diffusive process and splits the operators of mixing and burning \citep{pignatariNuGridStellarData2016}, so care has been taken that our post-processing models are converged with respect to spatial and temporal resolution.
\cite{issaImpact3DMacro2025} provides further explanation of all simulation details.

\subsection{Estimating Pre-Explosive Yields}\label{sec:estimate_production}

The post-processed models in this work are calculated only for the \oxygen[] shell without the rest of the star, including the merging \carbon[] shell.
Therefore, we must estimate the final yields if our mixing prescriptions were implemented into a full stellar evolution calculation.
The final ejected mass of a species $i$ is given by:
\[
    \mathrm{EM}_i = \mathrm{EM}^\mathrm{wind}_{i} + \mathrm{EM}^\mathrm{SN}_{i}
\]
where $\mathrm{EM}^\mathrm{wind}_{i}$ are the yields from winds throughout stellar evolution and $\mathrm{EM}^\mathrm{SN}_{i}$ are the ejected yields due to a supernova without explosive burning as described by \cite{ritterNuGridStellarData2018}. 
Winds are taken from the surface abundances and are not affected by the mixing conditions in the \oxygen[] shell, and are the same as \cite{ritterNuGridStellarData2018}, but the pre-explosive supernova yields are altered by the mixing conditions. 
Our assumption is that the ratio of the decayed mass fractions calculated in our models at $t=\unit{110}{\second}$ to those in the \cite{ritterNuGridStellarData2018} \oxygen[] shell at $\log_{10}(t - t_{\mathrm{end}})/\mathrm{yr} = -3.856$ (the corresponding time in the NuGrid model) remains the same for all subsequent evolution across the whole \oxygen[]-\carbon[] region. 
Since the convective turnover is shorter than the simulation time, this assumption is reasonable.
This allows us to estimate what the yields of an entire stellar model would be if our 3D-inspired mixing conditions were used, although this neglects feedback effects that cannot be captured without full stellar evolution calculations.
Therefore, these results are not meant as predictive yields, but rather an exploratory study of how mixing conditions may impact nucleosynthesis.

Following \cite{ritterNuGridStellarData2018}, we use the delayed explosion mass cut of \cite{fryerCOMPACTREMNANTMASS2012} to find $m=\unit{1.619}{\Msun}$ for our approximated yields:
\[
    \mathrm{EM}^\mathrm{SN}_{i} = \int_{\unit{1.619}{\Msun}}^{\unit{2.95}{\Msun}} X_i\cdot \mathrm{ratio}_i~ dm + \int_{\unit{2.95}{\Msun}}^{m_\tau} X_i~ dm
\]
where $X_i$ is the mass fraction of an element $i$, $\mathrm{ratio}_i$ is the ratio across the \oxygen[] shell of the decayed mass fractions for a model and \cite{ritterNuGridStellarData2018} for element $i$, $m_\tau$ is the extent of the star at collapse, and \unit{1.619{-}2.95}{\Msun} is the \oxygen[]-\carbon[] merger region above the mass-cut.
The mass fraction is calculated as the yield of a given species normalized to the total yield:
\begin{equation}\label{eq:Xf}
   X_i = \mathrm{EM}_i \div \sum_i^N{\mathrm{EM}}_i
\end{equation}
where $N$ is the total number of species.

The final results are calculated in square bracket notation against \cite{ritterNuGridStellarData2018} (indicated by the subscript $\mathrm{ref = R}$) and the solar measurements from \cite{asplundChemicalCompositionSun2009} ($\mathrm{ref = \odot}$):
\[
    [\mathrm{X/Y}]_{\mathrm{ref}} = \log_{10}(\mathrm{X/Y})_{\mathrm{norm}} - \log_{10}(\mathrm{X/Y})_{\mathrm{ref}}
\]
where $\mathrm{X}$ and $\mathrm{Y}$ are mass fractions calculated with \Eq{Xf}, and the subscript $\mathrm{norm}$ indicates what is being compared to the reference value.

\subsection{Estimating Reaction Contribution}

The reaction flux $f_{ij}$ is defined as the net flux of the reaction between species $i$ and $j$:
\[
    f_{ij} = \frac{X_i X_j}{A_i A_j} \rho N_A \Biggl(\langle \sigma v \rangle_{ij} - \langle \sigma v \rangle_{ji}\Biggr)
\]
where $X$ is the mass fraction, $A$ is the atomic mass, $\rho$ is the density, $N_A$ is Avogadro's
number, and $\langle \sigma v \rangle_{ij}$ is the reaction rate between species $i$ and $j$.

To estimate the total contribution of a reaction to the production and destruction of a species, we
calculate the average reaction flux $F_{ij}$ over mass and time:
\[
    F_{ij} = \frac{1}{\delta t (m_{\mathrm{top}} - m_{\mathrm{bot}})} \int_{t_c}^{t_c+\delta t} \int_{m_{\mathrm{bot}}}^{m_{\mathrm{top}}} |f_{ij}(m, t)| dm dt
\]
Using this, we can define what percentage a particular reaction contributes to the total flux for a species:
\begin{equation}
    \label{eq:contribution}
    \langle \mathrm{contribution} \rangle_{ij} = \frac{F_{ij}}{\sum_{j} F_{ij}}
\end{equation}
calculated at $t_\mathrm{c} = \unit{110}{\sec}$ and $\delta t = \unit{1}{sec}$. 
This provides quasi-equilibrium reaction fluxes at the end of the simulation.

\section{Results} \label{sec:results}

\subsection{Nucleosynthesis of the light odd-Z isotopes}\label{sec:lightoddZ}

The reference case is the nucleosynthesis of the light odd-Z isotopes in the \oxygen[] shell with
the MLT-based mixing efficiency profile with an ingestion rate of
$\unit{\natlog{4}{-3}}{\Msun\second^{-1}}$. \oxygen[16] produces \pt, \nt, and $\alpha$ by the main
O-burning reaction $\oxygen[16]+\oxygen[16]$, but the ingestion of \carbon[12] adds to the \pt, \nt,
and $\alpha$ release by $\carbon[12]+\carbon[12]$ and $\carbon[12]+\oxygen[16]$ reactions at
\oxygen[]-burning temperatures. The sudden increase of \pt, \nt, and $\alpha$ increases the capture
reactions $(\pt,\gamma)$, $(\nt,\gamma)$, $(\alpha,\gamma)$, $(\alpha,\pt)$, and $(\pt,\alpha)$
significantly which is one of the reasons why the light odd-Z elements are produced much more during
the merger than the first (cf.\ \Sect{methods_model}) convective \oxygen[] shell.

\begin{figure}
\includegraphics[width=\columnwidth]{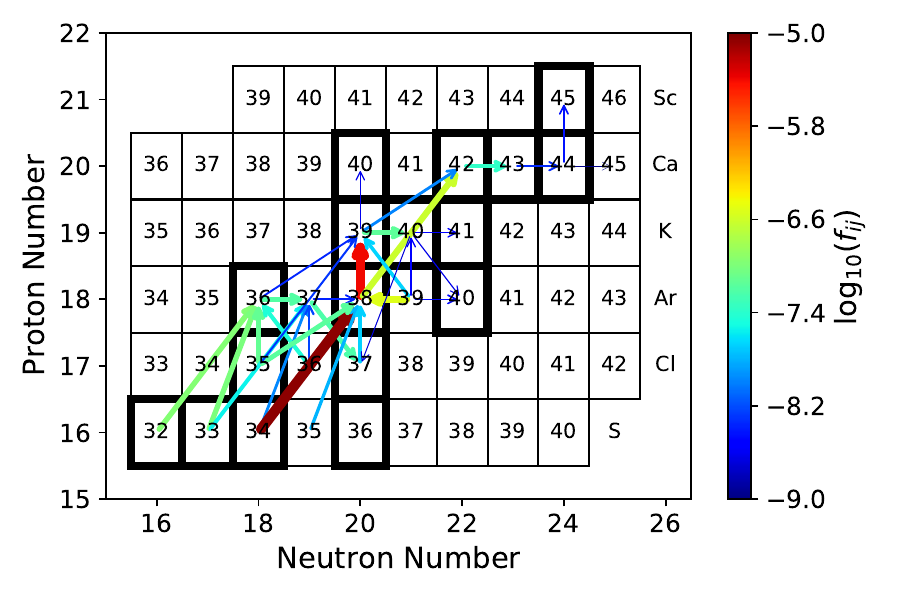}
\caption{Chart of reactions between isotopes at $m = \unit{1.65}{\Msun}$ [$T = \unit{2.231}{\Giga\K}$] for the MLT mixing case with an ingestion rate of $\unit{\natlog{4}{-3}}{\Msun\second^{-1}}$ at $t=\unit{110}{\second}$. Both arrow colour and size indicate $\log_{10}(f_{ij})$, and arrows point in the direction of the reaction.
\label{fig:reaction_chart}}
\end{figure}

\begin{figure}
\includegraphics[width=\columnwidth]{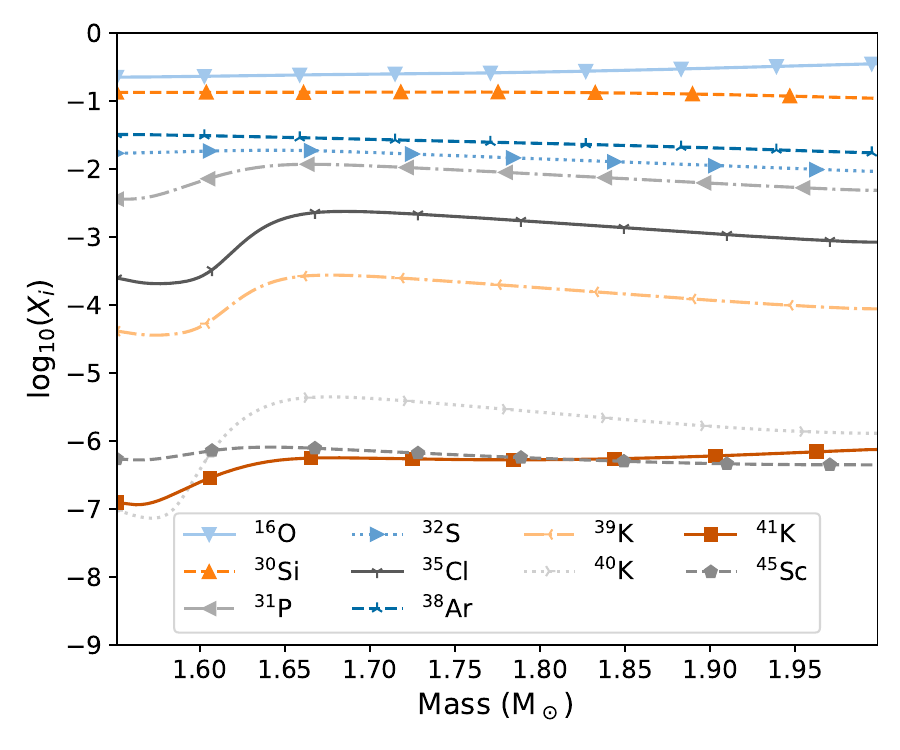}
\caption{Isotopic mass fractions at $t=\unit{110}{\second}$ without decays for the MLT mixing case with an ingestion rate of $\unit{\natlog{4}{-3}}{\Msun\second^{-1}}$.
\label{fig:isotopic_X}}
\end{figure}

The main reaction channels are shown in \Fig{reaction_chart}. Many additional reactions that are not
shown as the convective-reactive nature of the \oxygen[] shell means different locations in the
shell have varying contributions. \Fig{isotopic_X} shows the isotopic mass fractions across the
\oxygen[] shell, and as it shows the light odd-Z isotopes have a peak \unit{1.6{-}1.7}{\Msun}.

Important isotopes like \silicon[30], \sulfur[32,34], and \argon[38] are created in the first
convective \oxygen[] shell, and \chlorine[37], \argon[40], and \calcium[44] are created in the
\carbon[] shell and ingested during the merger. These pathways were determined by analyzing the
production of each of the species during the evolution of the \cite{ritterNuGridStellarData2018}
model and the fluxes in our post-processed models. The only stable \phosphorus[] isotope
\phosphorus[31] is produced by $\oxygen[16](\oxygen[16],\pt)\phosphorus[31]$. The stable \chlorine[]
isotope \chlorine[35] is produced by
$\silicon[30](\alpha,\gamma)\sulfur[34](\pt,\gamma)\chlorine[35]$,
$\sulfur[32](\alpha,\pt)\chlorine[35]$, and $\sulfur[35](\pt,\nt)\chlorine[35]$. \chlorine[37] is
both ingested from the \carbon[] shell and produced by $\sulfur[34](\pt,\alpha)\chlorine[37]$ and
$\chlorine[35](\nt,\gamma)\chlorine[36](\nt,\gamma)\chlorine[37]$. The principal stable isotope of
\potassium[], \potassium[39] is produced by $\argon[38](\pt,\gamma)\potassium[39]$, the long-lived
radioisotope \potassium[40] is produced by $\potassium[39](\nt,\gamma)\potassium[40]$, and the other
stable \potassium[] isotope \potassium[41] is produced by $\potassium[40](\nt,\gamma)\potassium[41]$
and $\argon[40](\pt,\gamma)\potassium[41]$. Finally, the only stable \scandium[] isotope
\scandium[45] is produced by
$\argon[38](\alpha,\gamma)\calcium[42](\nt,\gamma)\calcium[43](\nt,\gamma)\calcium[44](\pt,\gamma)\scandium[45]$.
However, our analysis shows that different reactions dominate the light odd-Z isotopic production depending on the mixing conditions as shown in Tables \ref{tab:reaction_contributions_MLT} and \ref{tab:reaction_contributions_PPM50}.
Additionally, the transient before the quasi-equilibrium at the end of the simulation will have different dominant reactions, but we assume that this is not important for the final yields.

\begin{table*}[ht]
\centering
\begin{tabular}{lccccccccccccc}
\toprule
Isotope & \oxygen[16](\oxygen[16],\pt) & ($\alpha$,$\gamma$) & ($\alpha$,\nt) & ($\alpha$,\pt) & (\pt,$\gamma$) & (\pt,\nt) & (\nt,$\gamma$) & ($\gamma$,\nt) & ($\gamma$,\pt) & (\nt,\pt) & (\nt,$\alpha$) & (\pt,$\alpha$) \\
\toprule
\phosphorus[31] & $+0.97$ &  &  & $-0.01$ & $+0.01$ &  &  &  &  &  &  &  \\
\chlorine[35] &  & $+0.01$ &  & $+0.18$ & $-0.57$ & $+0.22$ &  & $-0.01$ & $+0.01$ &  & &  \\
\chlorine[37] &  &  &  & $+0.23$ & $+0.07$ &  & $+0.02$ &  & $-0.45$ & $+0.22$ &  &  \\
\potassium[39] &  & $+0.02$ &  & $+0.04$ & $-0.71$ & $+0.14$ &  & $-0.05$ & $+0.01$ &  &  & $-0.02$ \\
\potassium[40] &  &  & $-0.09$ &  & $+0.03$ & $-0.05$ & $+0.64$ & $-0.09$ & $-0.04$ & $-0.05$ & &  \\
\potassium[41] &  & $+0.1$ &  & $-0.28$ & $+0.24$ &  & $+0.18$ &  & $-0.03$ & $-0.17$ &  & \\
\scandium[45] &  &  &  &  & $+0.49$ & $+0.01$ &  & $-0.07$ & $-0.42$ &  &  &  \\
\toprule
\end{tabular}
\caption{Reaction fluxes contributing to the light odd-Z isotopes using \Eq{contribution} for the MLT mixing case with an ingestion rate of $\unit{\natlog{4}{-3}}{\Msun\second^{-1}}$. Contributions starting with $+$ are productive (should be read as $i\rightarrow j$ e.g. \sulfur[34]$(\pt,\gamma)$\chlorine[37] with $+0.23$), those starting with $-$ are destructive (should be read as $i \leftarrow j$ e.g. \sulfur[34]$(\pt,\gamma)$\chlorine[35] with $-0.57$), and those left blank contribute less than $1\%$.}
\label{tab:reaction_contributions_MLT}
\end{table*}

\begin{table*}[ht]
\centering
\begin{tabular}{lccccccccccccc}
\toprule
Isotope & \oxygen[16](\oxygen[16],\pt) & ($\alpha$,$\gamma$) & ($\alpha$,\nt) & ($\alpha$,\pt) & (\pt,$\gamma$) & (\pt,\nt) & (\nt,$\gamma$) & ($\gamma$,\nt) & ($\gamma$,\pt) & (\nt,\pt) & ($\gamma$,$\alpha$) & (\nt,$\alpha$) & (\pt,$\alpha$) \\
\toprule
\phosphorus[31] & $+0.91$ &  &  & $-0.04$ & $+0.02$ &  &  &  & $-0.01$ &  &  & $-0.01$ \\
\chlorine[35] &  &  &  & $+0.05$ & $-0.9$ & $+0.01$ &  &  & $-0.03$ &  &  & $-0.01$ \\
\chlorine[37] &  &  &  & $-0.06$ & $+0.03$ &  &  &  & $-0.52$ & $+0.38$ &  &  \\
\potassium[39] &  &  &  & $+0.01$ & $-0.96$ &  &  &  & $-0.01$ &  &  & $-0.01$ \\
\potassium[40] &  &  & $-0.02$ & $-0.03$ & $+0.02$ & $-0.01$ & $+0.33$ & $-0.02$ & $-0.2$ & $-0.37$ &  &  \\
\potassium[41] &  & $+0.11$ &  & $-0.33$ & $+0.2$ &  & $+0.15$ &  & $-0.05$ & $-0.15$ & $+0.01$ &  \\
\scandium[45] &  &  &  & $+0.08$ & $+0.29$ &  & $+0.01$ & $-0.02$ & $-0.49$ & $+0.1$ &  &  \\
\toprule
\end{tabular}
\caption{Reaction fluxes contributing to the light odd-Z isotopes using \Eq{contribution} for the $50\times D_{\mathrm{3D{-}insp.}}$ mixing case with an ingestion rate of
$\unit{\natlog{4}{-3}}{\Msun\second^{-1}}$. Contributions starting with $+$ are productive (should be read as $i\rightarrow j$ e.g. \potassium[39]$(\nt,\gamma)$\potassium[40] with $+0.33$), those starting with $-$ are destructive (should be read as $i \leftarrow j$ e.g. \calcium[40]$(\nt,\pt)$\potassium[40] with $-0.37$), and those left blank contribute less than $1\%$.}
\label{tab:reaction_contributions_PPM50}
\end{table*}

\subsection{Impact of macro physics mixing on light odd-Z nucleosynthesis production}
\label{sec:potassium}

\begin{figure}
\includegraphics[width=\columnwidth]{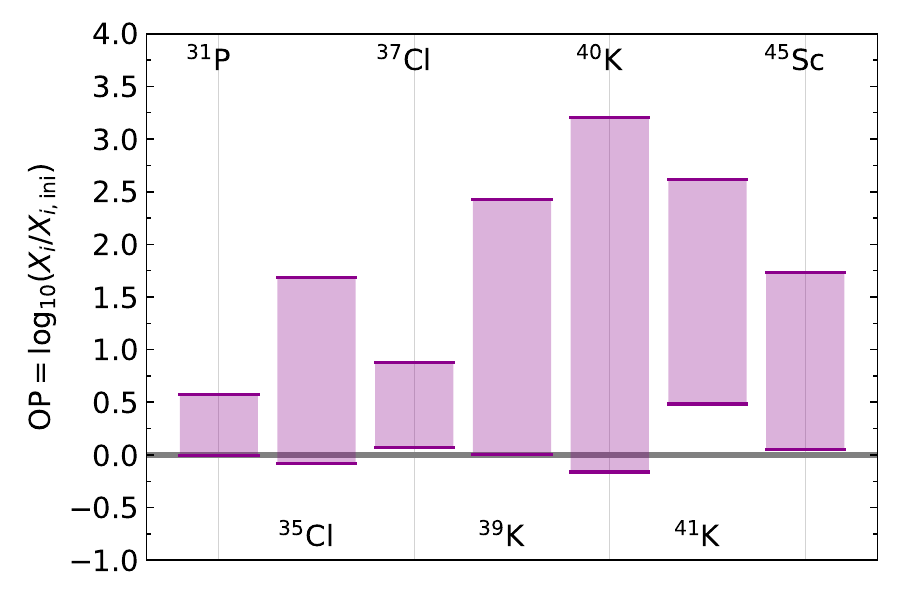}
\caption{Overproduction of the light odd-Z isotopes for all 24 mixing cases in the \oxygen[] shell at $\unit{110}{\second}$. The average spread $\OP_{\mathrm{max}} - \OP_{\mathrm{min}} = \unit{1.82}{\dex}$, and the line at $\OP=0$ is the initial amount.}
\label{fig:lightoddZ_barplot}
\end{figure}

\Fig{lightoddZ_barplot} shows the spread across all 24 mixing cases in overproduction in the \oxygen[] shell only at the end of the merger, $\OP = \log_{10}(X_i/X_{i,\mathrm{ini}})$, where $X_i$ is the final decayed, mass-averaged
mass fraction of an isotope and $X_{i,\mathrm{ini}}$ is the initial mass fraction at the start of the merger.
Over the range of mixing scenarios and entrainment rates, \phosphorus[31] has a spread $\OP_{\mathrm{max}} - \OP_{\mathrm{min}}$ of \unit{0.58}{\dex}, \chlorine[35] has \unit{1.76}{\dex}, \chlorine[37] has \unit{0.81}{\dex}, \potassium[39] has \unit{2.42}{\dex}, \potassium[40] has \unit{3.37}{\dex}, \potassium[41] has \unit{2.31}{\dex}, and \scandium[45] has \unit{1.68}{\dex} with an average spread across all light odd-Z isotopes of \unit{1.82}{\dex}. 
The \potassium[] isotopes are the most affected by the mixing conditions, and \potassium[40] has the largest spread of all isotopes. 
This underscores the importance of understanding the 3D macro physics from hydrodynamic simulations to reliably predict the final production during a merger.

\begin{figure}
\includegraphics[width=\columnwidth]{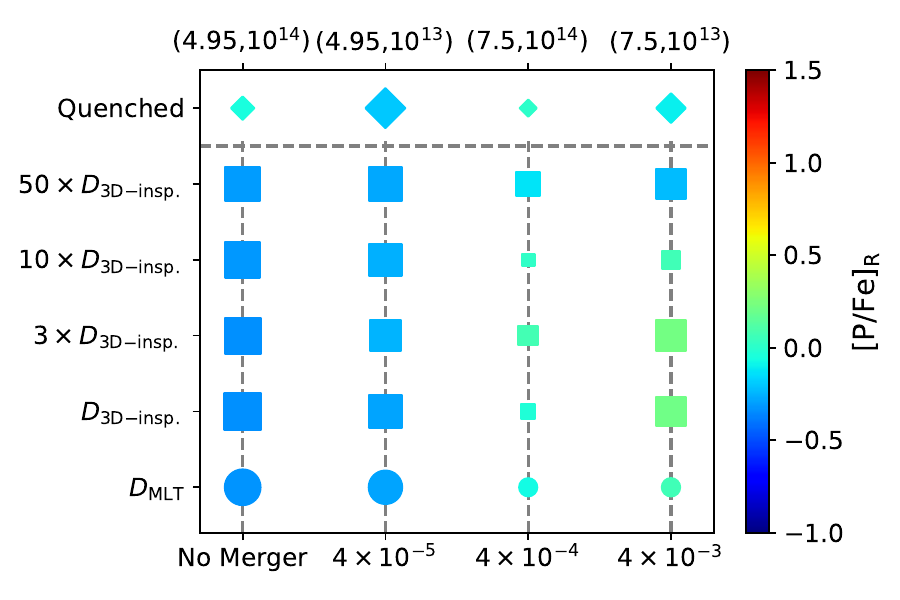}
\caption{The change in \dex{} for [\phosphorus[]/\iron[]] to the $\mzams=\unit{15}{\Msun}$
$Z=0.02$ model from \cite{ritterNuGridStellarData2018} as described in \Sect{estimate_production}.
The lower x-axis is the ingestion rate in $\Msun\second^{-1}$ for the MLT (circles) and 3D-inspired
mixing (squares) cases. The upper x-axis is the centre of the mixing efficiency dip in \Mm{} and the
extent of the dip in $\cm^{2}\second^{-1}$ for the quenched (diamonds) mixing cases. Size indicates
distance from $[\phosphorus[]/\iron[]]_{\mathrm{R}}=0$, and colour indicates the magnitude.
The \cite{ritterNuGridStellarData2018} full delayed pre-explosive yield $[\phosphorus[]/\iron[]]_\odot = \unit{2.13}{\dex}$, and the contributions just from winds are \unit{-0.04}{\dex}}. 
\label{fig:ratio_p}
\end{figure}

Figures \ref{fig:ratio_p}{--}\ref{fig:ratio_sc} show the modification to the pre-explosive yields of \cite{ritterNuGridStellarData2018} as described in \Sect{estimate_production}. 
The ratios [\phosphorus[]/\iron[]], [\chlorine[]/\iron[]], [\potassium[]/\iron[]], and [\scandium[]/\iron[]] are changed by $\unit{[-0.33,0.23]}{\dex}$, $\unit{[-0.84,0.64]}{\dex}$, $\unit{[-0.78,1.48]}{\dex}$, and $\unit{[-0.36,1.29]}{\dex}$ respectively.
The elemental abundances are most enhanced for the fastest ingestion rate of $\unit{\natlog{4}{-3}}{\Msun\second^{-1}}$ and for the mixing cases with a downturn in the mixing efficiency.
This enhancement is non-monotonic with increasing convective velocities for all elements
except \potassium[] which is indeed most enhanced for the $50\times D_{\mathrm{3D{-}insp.}}$ mixing case.

\begin{figure}
\includegraphics[width=\columnwidth]{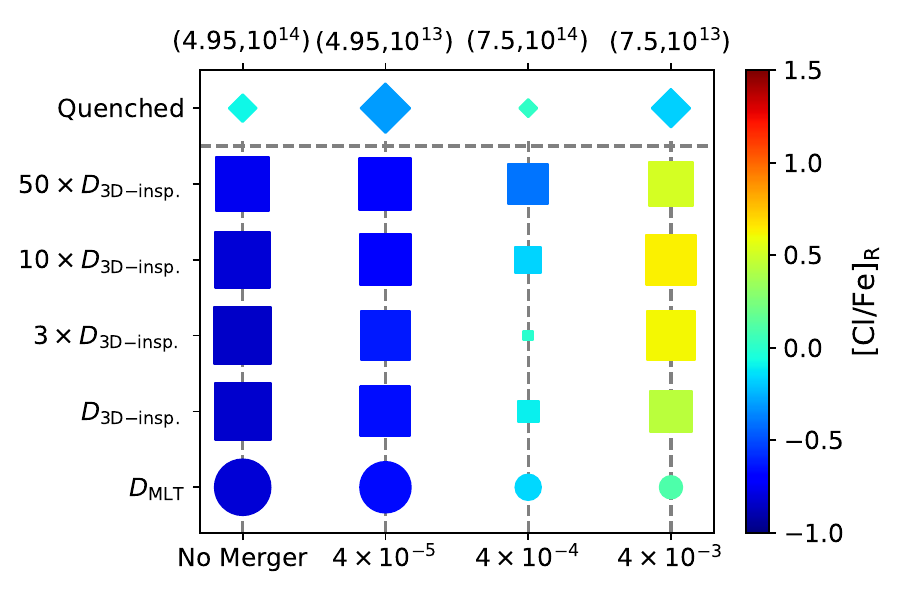}
\caption{The same as \Fig{ratio_p} but for $[\chlorine[]/\iron[]]_\mathrm{R}$. 
The \cite{ritterNuGridStellarData2018} full delayed pre-explosive yield $[\chlorine[]/\iron[]]_\odot = \unit{2.14}{\dex}$, and the contributions just from winds are \unit{0.21}{\dex}}.
\label{fig:ratio_cl}
\end{figure}

\begin{figure}
\includegraphics[width=\columnwidth]{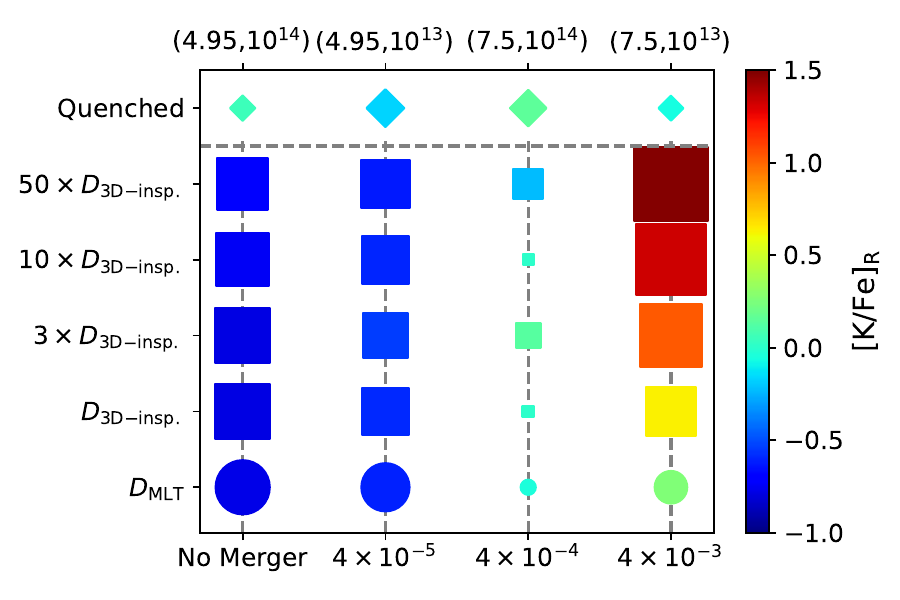}
\caption{The same as \Fig{ratio_p} but for $[\potassium[]/\iron[]]_\mathrm{R}$. 
The \cite{ritterNuGridStellarData2018} full delayed pre-explosive yield $[\potassium[]/\iron[]]_\odot = \unit{1.59}{\dex}$, and the contributions just from winds are \unit{-0.03}{\dex}}.
\label{fig:ratio_k}
\end{figure}

\begin{figure}
\includegraphics[width=\columnwidth]{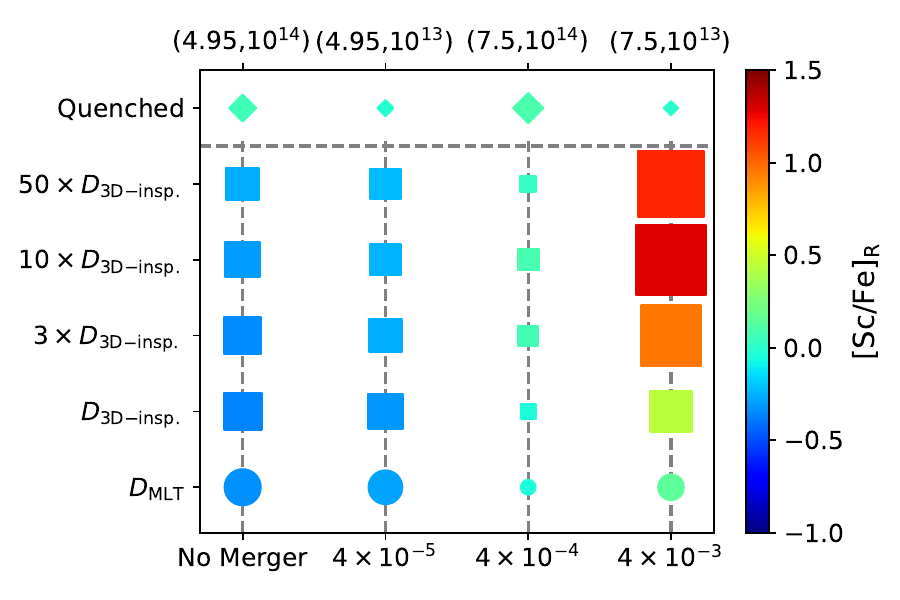}
\caption{The same as \Fig{ratio_p} but for $[\scandium[]/\iron[]]_\mathrm{R}$. The \cite{ritterNuGridStellarData2018} full delayed pre-explosive yield $[\scandium[]/\iron[]]_\odot = \unit{1.79}{\dex}$, and the contributions just from winds are \unit{0.06}{\dex}}.
\label{fig:ratio_sc}
\end{figure}

\begin{figure}
\includegraphics[width=\columnwidth]{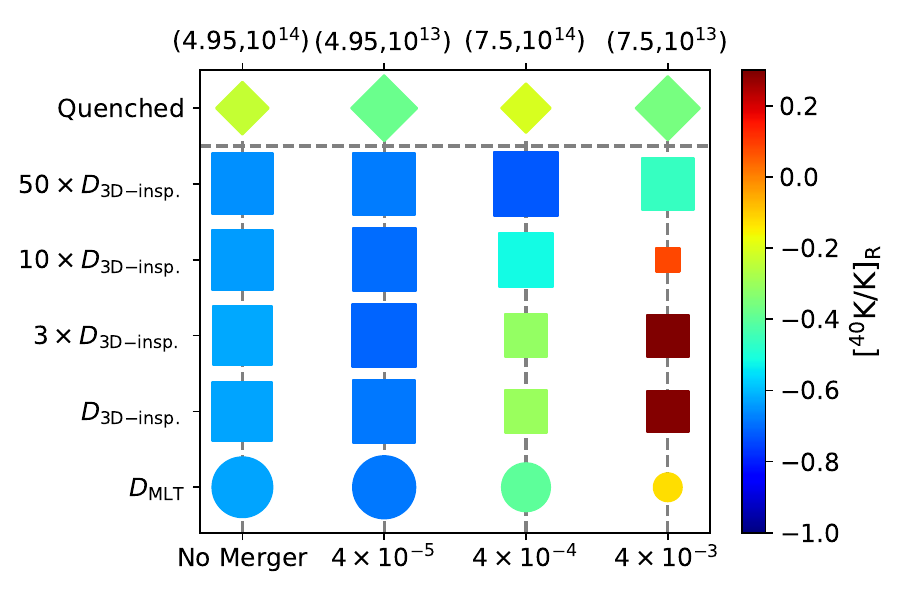}
\caption{The same as \Fig{ratio_p} but for $[\potassium[40]/\potassium[]]_\mathrm{R}$. 
The \cite{ritterNuGridStellarData2018} full delayed pre-explosive yield $[\potassium[40]/\potassium[]]_\odot = \unit{1.54}{\dex}$, and the contributions just from winds are \unit{-1.32}{\dex}.
Note that the colourbar is different from Figures \ref{fig:ratio_p}{--}\ref{fig:ratio_sc}.}
\label{fig:k40_k}
\end{figure}

Isotopes of the same element are not affected the same way by the mixing conditions (\Fig{lightoddZ_barplot}), and therefore elemental and isotopic enhancements are not necessarily aligned.
Elemental \potassium[] has the largest enhancement for the $50\times D_{\mathrm{3D{-}insp.}}$ mixing case (\Fig{ratio_k}), but the isotope \potassium[40] is most enhanced for the $1$ and $3\times D_{\mathrm{3D{-}insp.}}$ mixing cases (\Fig{k40_k}). 
Elemental K is primarily \potassium[39]. 
The enhancement of \potassium[40] in the lower mixing cases implies that the mixing conditions affect how these isotopes are produced relative to each other.

\begin{figure}
\includegraphics[width=\columnwidth]{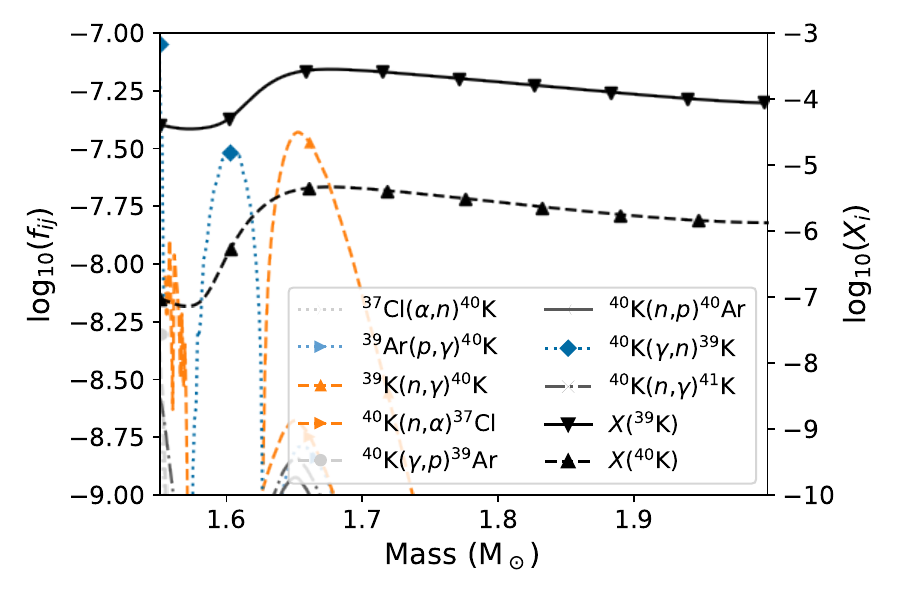}
\caption{Reaction fluxes $f_{ij}$ for \potassium[40] and mass fractions $X_i$ for \potassium[39,40] for the MLT mixing case with an ingestion rate of $\unit{\natlog{4}{-3}}{\Msun\second^{-1}}$ at $t=\unit{110}{\second}$. The direction of reactions $i
\rightarrow j$ are written as left to right in the legend. Note that the reaction direction notation as always left to right is different than Table \ref{tab:reaction_contributions_MLT}.}
\label{fig:k40_MLT}
\end{figure}

\begin{figure}
\includegraphics[width=\columnwidth]{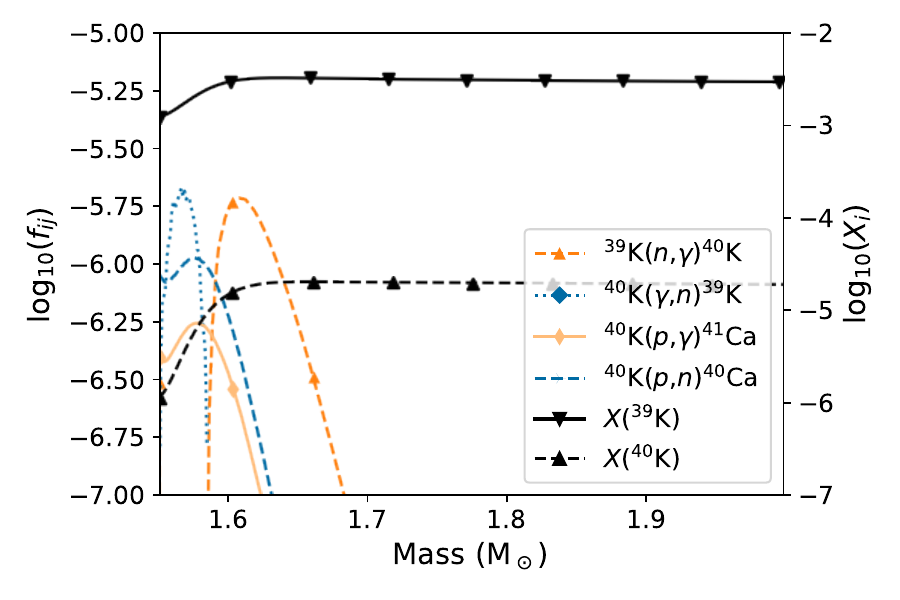}
\caption{The same as \Fig{k40_MLT} but for the $50\times D_{\mathrm{3D{-}insp.}}$ mixing case with an ingestion rate of $\unit{\natlog{4}{-3}}{\Msun\second^{-1}}$. Note that the reaction direction notation as always left to right is different than Table \ref{tab:reaction_contributions_PPM50}.}
\label{fig:k40_PPM50}
\end{figure}

Mixing conditions affect nucleosynthesis in a non-monotonic way because the \oxygen[] shell is a convective-reactive environment. 
As explored in \cite{issaImpact3DMacro2025}, when the mixing speeds change, so does the location of peak burning. 
For example, consider the reactions contributing to
the \potassium[40] mass fraction for the MLT mixing case in \Fig{k40_MLT} and the $50\times D_{\mathrm{3D{-}insp.}}$ mixing case in \Fig{k40_PPM50}.

Comparing Figures \ref{fig:k40_MLT} and \ref{fig:k40_PPM50} we see that for the faster mixing case (1) the reaction fluxes $\log_{10}(f_{ij})$ are stronger (2) the mass fractions $X_i$ are higher and have a different shape (3) the location of the strongest reactions are deeper in the shell (4) the dominant reactions are different.
In both cases, \potassium[40] is both produced and destroyed by $\potassium[39](\nt,\gamma)\potassium[40]$, but in the faster mixing case these reactions occur at hotter temperatures deeper in the shell and the destructive channels $\potassium[40](\pt,\nt)\calcium[40]$ and $\potassium[40](\pt,\gamma)\calcium[41]$ are much stronger. 
Note that \potassium[40] is being net destroyed in the faster mixing case, but still has a higher mass fraction because \potassium[39] is being produced even more.
This is evident when comparing to \Fig{k40_k} where the $50\times D_{\mathrm{3D{-}insp.}}$ mixing case has a lower [\potassium[40]/\potassium[]] ratio than the MLT case.
Since the shell is convective-reactive, the nucleosynthesis is sensitive to the ratio of the timescales for mixing and burning which changes the location where reaction peaks occur.
This behaviour is relevant for all nucleosynthesis in the \oxygen[] shell and is the cause of the spread in \Fig{lightoddZ_barplot} and elemental abundance enhancements in Figures \ref{fig:ratio_p}{--}\ref{fig:ratio_sc}.

\section{Heterogeneous \potassium[40] interstellar medium enrichment and impact for rocky planet heating}  \label{sec:k40_planets}

As discussed in the previous sections, all \potassium[] isotopes are efficiently produced in O-C shell mergers.
It is known that this could reconcile the [\potassium[]/\iron[]] observations in the Milky Way disk with GCE simulations
\citep[][]{ritterConvectivereactiveNucleosynthesisSc2018}.  \potassium[40] receives the biggest boost compared to the solar scaled initial abundances.
Let's consider as example the \unit{15}{\Msun} $Z=0.02$ model by
\cite{ritterConvectivereactiveNucleosynthesisSc2018}, affected by \oxygen[]-\carbon[] shell merger, and the analogous non-rotating model in the stellar set by \cite{limongi:18}.
The \potassium[39], \potassium[40] and \potassium[41] yields by the model with \oxygen[]-\carbon[] shell merger are \unit{\natlog{1.959}{-3}}{\Msun}, \unit{\natlog{1.165}{-4}}{\Msun} and \unit{\natlog{6.694}{-5}}{\Msun}, respectively.
The model without \oxygen[]-\carbon[] shell merger ejects \unit{\natlog{1.061}{-4}}{\Msun}, \unit{\natlog{6.071}{-7}}{\Msun} and \unit{\natlog{1.046}{-5}}{\Msun}, respectively. 
While \potassium[39] and \potassium[41] boosts are in the order of a factor of 10 or less \citep[][]{ritterConvectivereactiveNucleosynthesisSc2018}, the \potassium[40] increase in our 1D model with \oxygen[]-\carbon[] shell merger is almost $200$ times larger compared to a standard CCSN production
seen in the model by \cite{limongi:18}.
Additionally, we consider yields from our %least and 
most enhanced \potassium[40] production cases, %the $D_\mathrm{3D{-}insp.}$ no merger case and 
the $10\times D_{\mathrm{3D{-}insp.}}$ case with an ingestion rate of \unit{\natlog{4}{-3}}{\Msun}, which should be considered as an indicative upper limit in terms of \potassium[40] yields based on the current simulations. 
%The $D_\mathrm{3D{-}insp.}$ case ejects \unit{\natlog{3.434}{-4}}{\Msun}, \unit{\natlog{5.055}{-6}}{\Msun} and \unit{\natlog{2.101}{-5}}{\Msun} of \potassium[39], \potassium[40] and \potassium[41], respectively, and the $10\times D_{\mathrm{3D{-}insp.}}$ case ejects \unit{\natlog{3.992}{-2}}{\Msun}, \unit{\natlog{3.054}{-3}}{\Msun} and \unit{\natlog{9.857}{-4}}{\Msun}, respectively.
The $10\times D_{\mathrm{3D{-}insp.}}$ case ejects \unit{\natlog{3.992}{-2}}{\Msun}, \unit{\natlog{3.054}{-3}}{\Msun} and \unit{\natlog{9.857}{-4}}{\Msun} of \potassium[39], \potassium[40] and \potassium[41], respectively.

\begin{figure}
\includegraphics[width=\columnwidth]{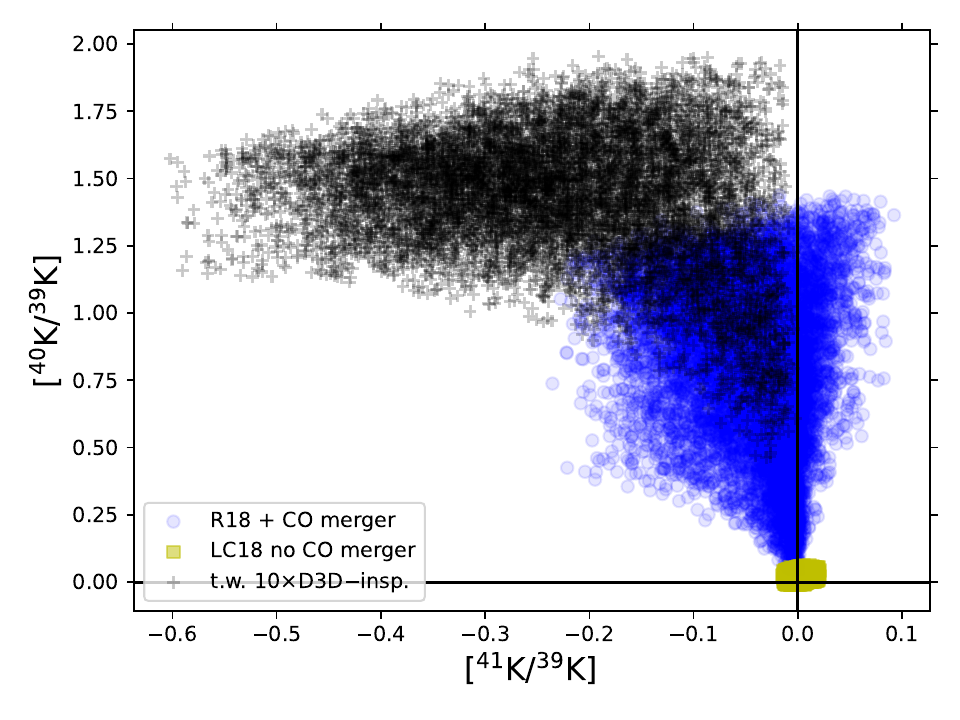}
\caption{Predicted dispersion of K isotopic ratios in the interstellar medium (ISM) resulting from
pollution by a single CCSN explosion, based on stellar yields from a \unit{15}{\Msun} star at solar
metallicity with \oxygen[]-\carbon[] shell merger \citep[R18,][]{ritterConvectivereactiveNucleosynthesisSc2018} and
without \citep[LC18][]{limongi:18}, as well as shell-merger only predictions from this work (t.w.)
for %the $D_\mathrm{3D{-}insp.}$ no merger case and 
the $10\times D_{\mathrm{3D{-}insp.}}$ mixing case with an ingestion rate of \unit{\natlog{4}{-3}}{\Msun}. 
Isotopic ratios are expressed in bracket notation, reported as
logarithms base 10, and normalized to solar values.}
\label{fig:k_ISM}
\end{figure}

Although this needs to be investigated further, it is likely that the occurrence of \oxygen[]-\carbon[] shell mergers and the associated strong boost in \potassium[40] takes place in a significant fraction of but not all massive stars \citep[][]{ritterConvectivereactiveNucleosynthesisSc2018}.
When it happens, this is likely with variable strength. 
To some degree the latter is reflected by the different assumed entrainment rates. 
Such efficient, yet variable production can introduce a heterogeneous signature of \potassium[40] in the local ISM once the CCSN explode, in addition to other \potassium[] sources in GCE. %, particularly for \potassium[40]. 
In \Fig{k_ISM}, we show the dispersion of K isotopic ratios in the ISM resulting
from pollution by two  single CCSN explosions, using the yields just provided and assuming a solar-like ISM composition. 
For each CCSN pollution scenario, we generated a random distribution of 10,000 points, considering a factor of 3 uncertainty in the K isotopic yields and a range of dilutions from \unit{\natlog{2}{3}}{\Msun} \citep[lower limit derived for a low-energy CCSN,][]{ji:20} to a typical value of $\unit{10^5}{\Msun}$ of ISM material. 
The obtained \potassium[41]/\potassium[39] dispersions are less than 10\% using the model without O-C shell merger and up to about a factor of two for the model with O-C shell merger by \cite{ritterConvectivereactiveNucleosynthesisSc2018}. 
Similar variations could also be expected in the solar neighborhood due to GCE effects, such as different metal enrichment
histories in the disk \citep[e.g.,][and references therein]{hegedus:25} and stellar migration
\citep[e.g.,][]{kubryk:15}. 
If instead we consider the $10\times D_{\mathrm{3D{-}insp.}}$ mixing case presented in the previous sections, we obtain a variation up to a factor of 4 from the solar \potassium[41]/\potassium[39] ratio.

On the other hand, local ISM pollution from the CCSN model by 
\cite{ritterConvectivereactiveNucleosynthesisSc2018} would generate a \potassium[40]/\potassium[39]
dispersion up to a factor of 20. Likewise, the $10\times D_{\mathrm{3D{-}insp.}}$ mixing case from
this work would produce an %similar 
even larger dispersion, up to about a factor of $100$, while the model without O-C
shell merger would show again an insignificant dispersion. Note, that we are showing in this way
only the $10\times D_{\mathrm{3D{-}insp.}}$ mixing case, which is the one with the largest
\potassium[40] production but not the one with the largest \potassium[40]/\potassium[39] ratio
(\Fig{k40_k}). The latter is obtained for the $1$ and $3\times D_{\mathrm{3D{-}insp.}}$ mixing cases.

The showcased \oxygen[]-\carbon[] shell merger cases would introduce a strongly heterogeneous signature in the GCE \potassium[] abundances of the ISM, locally decoupling the abundance of \potassium[40] from the other \potassium[] isotopes obtained from GCE.
Consequently, the Sun cannot be used as a robust reference for the \potassium[40]/\potassium[] ratio for stars in the solar neighborhood. 
Instead, our work shows that solar ``twins" with otherwise similar metallicity and \potassium[] abundance could have significantly different initial \potassium[40] abundance compared to the solar system.
 
The \potassium[40] isotopic effect described here has important thermal histories for rocky exoplanets. Energy produced from natural radioactive decay and the leftover primordial heat from planet formation powers the (internal) geodynamics of a rocky planet like Earth.
Rocky planet mantle temperatures govern, for example, expressed in crustal compositions, style of \citep[][]{rupkeSerpentineSubductionZone2004} as well as tectonic regime \citep[e.g.,][]{oneill:16,brown:20} 

Thermal evolution models offer a unique approach to understanding the evolution of temperatures
within rocky planetary interiors (Earth, Venus, Mercury, Mars, large asteroids, icy moons,
trans-neptunian objects, and 'terrestrial-type' exoplanets). These models are used to constrain
their volcanism and surface geology. Subordinate lithophile elements such as \potassium[] and \chlorine[] also help
stabilize minerals that control bulk transport properties, and/or volatile transport in rocky crusts
and mantles \citep[][]{fei:17, nishi:14, pamato:15, xu:08} %[REFS 53, 149, 158, 240]. 
A rocky planet's internal heat production is a function of potential and kinetic energy released
during accretion and core formation, and radioactive decay \citep[cf.\ ][]{labrosse:07} %[REF 106] 
thus, mantle heat production depends on planet(star) age. A planet's ability to sustain a mantle
convective regime and crustal processes changes considerably with its evolving thermal profile due
to cooling over geologic timescales of billions of years \citep[][]{sleep:00}. %[REFS 151, 194, 219]. 
Here, we have shown that across different extrasolar systems we can expect the starting inventories
of these nuclides to vary in abundance, including in their isotopes such as \potassium[40] even if
the elemental abundances may be similar to Solar. Despite the long-suspected (but hitherto,
unconstrained) variability, the majority of exoplanet studies up to now have applied Solar values
for radionuclide (U, Th, K) concentrations, often without considering the decay after planet
formation \citep[][]{frank:14}.   

\begin{figure}
\includegraphics[width=\columnwidth]{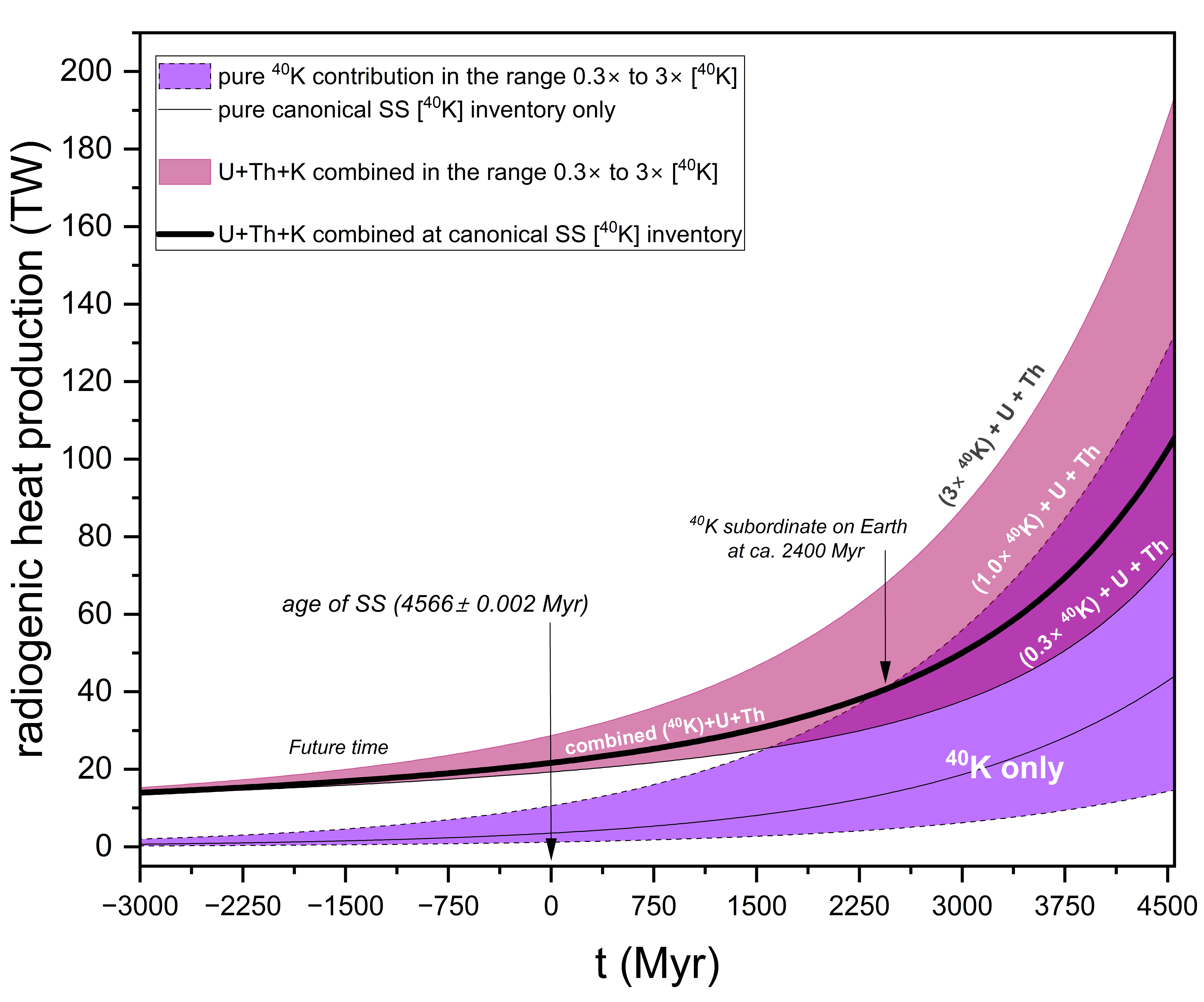}
\caption{The evolution over time of the radioactive heat production is shown for an Earth-like planet, where \potassium[40] abundances are increased and decreased by a factor of $3$ compared to the typical Earth values. The combined total heat productions by \potassium[40], \thorium[232], \uranium[235], and \uranium[238] are also shown. Time on the x-axis is in \Myr{} before present and so moves from right to left. {Negative values on the x-axis refer to future time.}}
\label{fig:k40radio}
\end{figure}

As an example, we illustrate this in \Fig{k40radio}, which shows the heating rate over time for Earth. 
To mimic the \potassium[40] variations observed in \Fig{k_ISM}, we multiply and divide the current abundance of \potassium[40] in Earth's silicate mass \citep[\unit{\natlog{3.04}{-8}}{kg/kg}, e.g.][and references therein]{frank:14} by a factor of $3$. 
Although similar terawatt (TW) values are obtained today by varying the \potassium[40] abundance, the planet's heating history would clearly be significantly different.

We will explore in detail elsewhere the implications of this for different types of rocky planets, considering more detailed pollution scenarios. 
Indeed, it is unclear whether the Solar System has a \potassium[40]/\potassium[] ratio consistent with GCE or if it reflects local pollution from CCSNe. 
Another interesting aspect, beyond the scope of this analysis, is whether we should expect local variations in the other radioactive sources, \uranium[] and \thorium[]. 
The current GCE of \thorium[] in the Milky Way disk is under debate \citep[e.g.,][]{botelho:19, mishenina:22}, while measuring \uranium[] in stellar spectra is even more challenging. Since \rpr{} sources are expected to be much rarer than CCSNe \citep[e.g.,][]{sneden:08, cowan:21}, heterogeneous signatures from local \uranium[235,238] and \thorium[232] pollution are likely to be extremely rare and possibly directly associated with the stellar \thorium[] abundance. 
However, the relevance of such heterogeneity affecting also the \rpr{} actinide abundances cannot be ruled out. 
It is worth noting that a significant \plutonium[244] abundance enhancement, where \plutonium[244] is a short-lived radioactive isotope also made by the \rprn, can be inferred for the Early Solar System compared to measurements representative of the present ISM, derived from deep-sea crust and other planetary materials \citep[e.g.,][]{hotokezaka:15, wallner:21, wehmeyer:23, wang:23, bishop:25}.

\section{Discussions and Conclusions} \label{sec:discussion} 

As shown in this analysis, \oxygen[]-\carbon[] shell mergers significantly modify the pre-explosive yields of the light odd-Z elements, particularly elemental \potassium[] which can be enhanced by up to $38\times$ compared to \cite{ritterNuGridStellarData2018}. 
These results reflect the systematic uncertainty in 1D stellar models due to the treatment of mixing, and highlight the potential instrinsic variability in massive stars.
This indicates the importance for 3D hydrodynamic simulations to constrain the mixing conditions during these events to improve predictions.
We have previously shown that mixing scenario changes which reaction rates are correlated with isotopic production \citep{issaImpact3DMacro2025}, so with realistic 3D mixing conditions the next critical step is to explore the uncertainties in these rates.
Observations can provide constraints, such as those that report a low dispersion in [\potassium[]/\iron[]] for metal-poor stars \citep[e.g.,][]{andrievskyNonLTEAbundancesMg2010,cohenNormalOutlyingPopulations2013,roedererSearchStarsVery2014,ishigakiPotassiumAbundancesExtremely2025}.
This could indicate that mixing conditions are more uniform in real stars than the range explored here, however GCE calculations that include a realistic range of shell merger conditions are needed to confirm this.
\cite{ritterConvectivereactiveNucleosynthesisSc2018} show that \oxygen[]-\carbon[] shell mergers are necessary to explain the light odd-Z elements in GCE, but \cite{robertiOccurrenceImpactCarbonOxygen2025} show that this could also lead to models producing more than observations. 
Future 3D modelling will be needed to define the correct mixing conditions, possibly yielding K abundances compatible with both stellar archaeology and GCE.
Finally, because the \oxygen[] shell is convective-reactive, the variations in yields due to mixing may not be the same for models with different temperature profiles or shell sizes.
Therefore, the results presented here cannot be extrapolated to other stellar models without further study.

Mixing conditions are set by the 3D macro physics of convection of the O-C shell merger and strongly affects the production of \potassium[40], which is a long-lived radionuclide with a half-life of $t_{1/2}=\unit{1.248}{\Giga\mathrm{yr}}$ that contributes to radiogenic heating in planets \citep{frankRadiogenicHeatingEvolution2014, oneillEffectGalacticChemical2020}. 
Not only is the elemental \potassium[] enhanced, but so is [\potassium[40]/\iron[]] up to $26\times$ compared to \cite{ritterNuGridStellarData2018} result for the $10\times D_{\mathrm{3D{-}insp.}}$ mixing case with an ingestion rate of $\unit{\natlog{4}{-3}}{\Msun\second^{-1}}$ (using square bracket notation adding the results of \Fig{ratio_k} and \Fig{k40_k}). 
This would mean that the \potassium[40] contributed to GCE is significantly enhanced. As we showed in \Fig{k_ISM}, there is a large variation in the \potassium[40]/\potassium[39] ratio which tracks the \potassium[40]/\potassium[] ratio as \potassium[39] is the dominant isotope of \potassium[]. 
This combined with the fact that the mixing conditions do not non-monotonically affect the production of \potassium[39] and \potassium[40] means that the \potassium[40]/\potassium[39] ratio can vary significantly between stars with otherwise similar metallicity and elemental \potassium[] observations, which brings into question the validity of deriving \potassium[40] from an elemental abundance measurement of the host star. 
Furthermore, we have shown that the production of \potassium[40] in \oxygen[]-\carbon[] shell mergers is so high that it could potentially introduce heterogeneous signatures in the local ISM.
With the reference CCSN model considered in this work, we find that \potassium[40] could be enhanced by up to a factor of 20 compared to its background GCE abundance, decoupling the \potassium[40] abundance from the observable \potassium[] abundances resulting from GCE. 
With the most \potassium[40]-rich mixing model discussed in this work, we find possible heterogeneous signatures affecting all \potassium[] isotopes, with a \potassium[40]-enhancement up to a factor of a hundred.

An enhanced \potassium[40] would have a significant effect on the history of radiogenic heat production in the
mantle of an Earth-like planet. A higher initial inventory of \potassium[40] raises the radiogenic heat
budget and sustains vigorous mantle convection and melt production for longer, which in turn
prolongs outgassing and delays tectonic/volcanic shutdown. Thermal-chemical evolution models show
that increased internal heating can keep silicate planets volcanically and climatically active for
Gyr timescales beyond nominal cases by maintaining continuous outgassed supply to the atmosphere and
postponing secular cooling \citep{FrankMeyerMojzsis2014}. In our context, a
$\sim 3\times$ enhancement in \potassium[40] plausibly extends the duration of high outgassing and
temperate surface conditions into later epochs relative to an Earth twin, while a $\sim10\times$
case would push the system toward a long-lived super-volcanic regime (still buffered by the
carbon-silicate feedback), thereby broadening the temporal window for habitability via sustained
greenhouse forcing and volatile cycling \citep{FoleySmye2018}.

%\paragraph{Diagnosing super-volcanic worlds via temporal spectral variability.}
An elevated \potassium[40] budget implies persistently high volcanic fluxes and a greater frequency of
large eruptions. Time-variable outgassing (e.g., SO$_2$ pulses that rapidly convert to sulfate
aerosols) imprints may be detectable, \emph{time-dependent} changes in atmospheric composition—most notably
in O$_3$ and H$_2$O bands—on Earth-like exoplanets \citep{OstbergEtAl2023}. Multi-epoch direct
imaging/spectroscopy with future facilities such as the Habitable Worlds Observatory may therefore have the potential to
discriminate a super-volcanic, high-\potassium[40] Earth-analog from a more quiescent twin \citep{LigginsEtAl2022,OstbergEtAl2023}.

Although our results demonstrate that mixing conditions strongly influence the production of the light odd-Z elements, they do not yet provide a complete picture. 
Our analysis examines a single stellar model at $Z=0.02$. The \potassium[39-41] mass fractions can vary by factors of $3$, $16$, and $38$ respectively in the merged \oxygen[]-\carbon[] shells listed in Table \ref{tab:merger_properties}. 
Model differences are not as impactful as mixing conditions, however, as \Sect{potassium} shows that final production of \potassium[39], \potassium[40], and \potassium[41] in the \oxygen[] shell can vary by factors of $263$, $2340$, and $204$ respectively across all mixing cases.

Our results in Figures \ref{fig:ratio_p}{--}\ref{fig:k40_k} omit the subsequent core-collapse
supernova and any explosive nucleosynthesis that may occur, and uses a first-order approximation to
estimate the final yields. The post-processing approach does not properly account for the evolution
of the merger in the 1D model, as temperature and density evolve and we do not simulate the mixing
of material in the \oxygen[] shell up into the merged \carbon[] shell. Further, it is not clear what
the length of the merger would be in 3D nor its behaviour for energy feedback in response to the
burning of ingested fuel, such as the recently identified SPAr process
\citep{robertiSPArProcessProton2025}. While these approximations limit presently the accuracy of
specific quantitative predictions, the result presented here, that 3D macro physics mixing conditions
strongly affect the nucleosynthesis, is robust. We also demonstrated that  the 1D mixing
prescriptions used here do not capture the full range of possible mixing behaviours that may occur in 3D.

Additionally, one would expect the MLT \unit{\natlog{4}{-3}}{\Msun \second^{-1}} mixing case to
match the \cite{ritterNuGridStellarData2018} model most closely in Figures
\ref{fig:ratio_p}{--}\ref{fig:k40_k}. The difference results from a combination of the above factors
and possibly also that in our targeted, zoom-in post-processing we took great care that  temporal
and spatial resolution is sufficient to resolve the convective-reactive nucleosynthesis, which cannot
be expected from survey stellar evolution models as those by \cite{ritterNuGridStellarData2018}.
This highlights the difficulty of capturing convective-reactive events in 1D stellar evolution
models and the need for 3D hydrodynamic simulations to inform 1D models.

The highest priority, therefore, is to investigate the macro physics of \oxygen[]-\carbon[] shell
mergers with fully 3D hydrodynamic simulations. Such simulations must quantify the mixing, identify
feedback mechanisms, and establish when and how often mergers occur. Ultimately, full
stellar-evolution calculations with mixing prescriptions calibrated to these 3D simulations are
needed to assess the impact of mixing conditions on pre-supernova and ejecta yields, and inform
planetary science studies of the impact of \potassium[40] on planetary thermal histories.

\begin{acknowledgments}
FH is supported by a Natural Sciences and Engineering Research Council of Canada (NSERC) Discovery
Grant and acknowledges support from the NSERC award SAPPJ-797 2021-00032 $\emph{Nuclear physics of
the dynamic origin of the elements}$. We acknowledge support from the ChETEC-INFRA project funded by
the European Union's Horizon 2020 Research and Innovation programme (Grant Agreement No 101008324).
This research has used the Astrohub online virtual research environment
(\url{https://astrohub.uvic.ca}), developed and operated by the Computational Stellar Astrophysics
group at the University of Victoria and hosted on the Digital Research Alliance of Canada Arbutus
Cloud at the University of Victoria. This work benefited from interactions and workshops
co-organized by The Center for Nuclear astrophysics Across Messengers (CeNAM) which is supported by
the U.S. Department of Energy, Office of Science, Office of Nuclear Physics, under Award Number
DE-SC0023128. MP acknowledges the support to NuGrid from the ``Lendület-2023'' Program of the
Hungarian Academy of Sciences (LP2023-10, Hungary), the ERC Consolidator Grant funding scheme
(Project RADIOSTAR, G.A. n. 724560, Hungary), the ChETEC COST Action (CA16117), supported by the
European Cooperation in Science and Technology, and the IReNA network supported by NSF AccelNet
(Grant No. OISE-1927130). MP also thanks the support from NKFI via K-project 138031 (Hungary). SJM
and MP acknowledge funding from the ERC Horizon Europe funding programme in support of the Synergy
Grant - Geoastronomy, grant agreement number 101166936. SJM also thanks the BGI/University of
Bayreuth in Germany for hosting the Geoastronomy Research Group.
\end{acknowledgments}

\begin{contribution}
JI performed the simulations, analyzed the results, and wrote
the manuscript. FH provided guidance and support for the project and edited the manuscript. MP
calculated the impact of heterogeneous mixing in the ISM of K isotopes, participating in data
analysis and in writing the manuscript. SM computed the differential heat production of variable
starting \potassium[40] abundances, participating in data analysis, and in writing the manuscript.
\end{contribution}

\software{The data and analysis tools are available on Zenodo under an open-source Creative Commons Attribution license: \dataset[doi:10.5281/zenodo.17576026]{https://doi.org/10.5281/zenodo.17576026}. They are also available for download from the NuGrid collaboration website: \url{download.nugridstars.org}.}

\bibliography{paper}{}
\bibliographystyle{aasjournalv7}

\end{document}

%% file: units.tex
\newcommand{\numberspace}{\ensuremath{\;}}

\newcommand{\unitstyle}[1]{\ensuremath{\mathrm{#1}}}
\newcommand{\hour}{\unitstyle{h}}

\newcommand{\natlog}[2]{\ensuremath{#1\times 10^{#2}}} % a*10^b   

\newcommand{\centi}{\unitstyle{c}}

\newcommand{\Mega}{\unitstyle{M}}
\newcommand{\Giga}{\unitstyle{G}}

\newcommand{\meter}{\unitstyle{m}}

\newcommand{\second}{\unitstyle{s}}
\newcommand{\Kelvin}{\unitstyle{K}}
\newcommand{\K}{\Kelvin}  %degrees Kelvin

\newcommand{\cm}{\centi\meter}

\newcommand{\Msun}{\ensuremath{\unitstyle{M}_\odot}}

\newcommand{\Myr}{\Mega\yr}
\newcommand{\Gyr}{\Giga\yr}

\newcommand{\yr}{\unitstyle{yr}}        %year
\newcommand{\Mm}{\Mega\meter}   %megameters
\newcommand{\OP}{\ensuremath{\unitstyle{OP}}} % overproduction

\newcommand{\unit}[2]{\ensuremath{#1\numberspace\mathrm{#2}}}

%% file: nuclides.tex
\newcommand{\nuclei}[2]{\ensuremath{\mathrm{^{#1}#2}}}

\newcommand{\neutron}{\ensuremath{n}}
\newcommand{\nt}{\neutron}
\newcommand{\proton}{\ensuremath{p}}
\newcommand{\pt}{\proton}

\newcommand{\carbon}[1][12]{\nuclei{#1}{C}}

\newcommand{\oxygen}[1][16]{\nuclei{#1}{O}}

\newcommand{\magnesium}[1][24]{\nuclei{#1}{Mg}}

\newcommand{\silicon}[1][28]{\nuclei{#1}{Si}}
\newcommand{\phosphorus}[1][31]{\nuclei{#1}{P}}
\newcommand{\sulfur}[1][32]{\nuclei{#1}{S}}
\newcommand{\chlorine}[1][35]{\nuclei{#1}{Cl}}
\newcommand{\argon}[1][36]{\nuclei{#1}{Ar}}
\newcommand{\potassium}[1][39]{\nuclei{#1}{K}}
\newcommand{\calcium}[1][40]{\nuclei{#1}{Ca}}
\newcommand{\scandium}[1][45]{\nuclei{#1}{Sc}}

\newcommand{\iron}[1][56]{\nuclei{#1}{Fe}}

\newcommand{\thorium}[1][232]{\nuclei{#1}{Th}}

\newcommand{\uranium}[1][238]{\nuclei{#1}{U}}

\newcommand{\plutonium}[1][244]{\nuclei{#1}{Pu}}

%% file: formatting.tex
\newcommand{\Tabff}[1]{{\ref{tab:#1}}}
\newcommand{\Tab}[1]{{Table~\Tabff{#1}}}

\newcommand{\pan}[1]{{\textit{#1}}}

\newcommand{\FIGFF}[2]{{\ref{fig:#2}\pan{#1}}}

\newcommand{\FIG}[2]{{Fig.~\FIGFF{#1}{#2}}}
\newcommand{\Fig}[1]{{\FIG{}{#1}}}

\newcommand{\Sectff}[1]{{\ref{sec:#1}}}
\newcommand{\Sect}[1]{{\S\Sectff{#1}}}

\newcommand{\Eqref}[1]{{\ref{eq:#1}}}
\newcommand{\Eqff}[1]{{(\Eqref{#1})}}

\newcommand{\Eq}[1]{{Eq.~\Eqff{#1}}}

\newcommand{\isofont}[1]{{\mathrm{#1}}}
\newcommand{\isomass}[1]{{\ensuremath{\isofont{^{#1}}}}}
\newcommand{\isocharge}[1]{{\ensuremath{\isofont{_{#1}}}}}
\newcommand{\isotope}[3]{{\ensuremath{\isocharge{#1}\isomass{#2}\isofont{#3}}}}

\newcommand{\I}[2]{{\isotope{}{#1}{#2}}}

\newcommand{\Ep}[1]{{\ensuremath{10^{#1}}}}
\newcommand{\E}[1]{{\ensuremath{\powersep\Ep{#1}}}}

\newcommand{\powersep}{\times}

\newcommand{\mem}[1]{\ensuremath{\mathrm{ #1}}}

%% file: derivatives.tex
\newcommand{\D}{{\mathrm d}}

%% file: concepts.tex
\newcommand{\rpr}{\mbox{$r$-process}}
\newcommand{\rprn}{\mbox{$r$ process}}
\newcommand{\ppr}{\mbox{$p$-process}}

\newcommand{\pnucn}{\mbox{$p$ nuclei}}

%% file: code.tex
\newcommand{\code}[1]{\texttt{#1}}

\newcommand{\mesa}{\code{MESA}}
\newcommand{\MESA}{\mesa}

\newcommand{\mppnp}{\code{mppnp}} % multi-zone post-processing network parallel

%% file: abbrev.tex
\newcommand{\n}{\ensuremath{\mem{n}}}

\newcommand{\p}{\ensuremath{\mem{p}}}
\newcommand{\e}{\ensuremath{\mem{e}}}

\newcommand{\h}{\ensuremath{\mem{H}}}

\newcommand{\nine}{\ensuremath{^{59}\mem{Ni}}}

\newcommand{\dex}{\ensuremath{\, \mathrm{dex}}} 
\newcommand{\mzams}{\ensuremath{M_{\rm ZAMS}}}

\newcommand{\beq}{\begin{equation}}
\newcommand{\beqa}{\begin{eqnarray}}
\newcommand{\eeq}{\end{equation}}
\newcommand{\eeqa}{\end{eqnarray}}
\newcommand{\bedis}{\begin{displaymath}}
\newcommand{\edis}{\end{displaymath}}